\documentclass[reprint,superscriptaddress,amsmath,amssymb,aps,pra]{revtex4-1}

\usepackage{times}
\usepackage{graphicx}
\usepackage{dcolumn}
\usepackage{bm}
\usepackage{braket}
\usepackage{mathtools}
\usepackage{dsfont}
\usepackage{url}
\usepackage{amsmath}
\usepackage{booktabs}
\usepackage{xcolor}
\usepackage{float}
\usepackage{url}

\graphicspath{{./Figures/}}

\begin{document} 

\title{Resource-efficient encoding algorithm for variational bosonic quantum simulations}
\author{Marco Majland}
\affiliation{Department of Physics and Astronomy, Aarhus University, Ny Munkegade 120, 8000 Aarhus C, Denmark}
\email{majland@phys.au.dk}
\author{Nikolaj Thomas Zinner}
\affiliation{Department of Physics and Astronomy, Aarhus University, Ny Munkegade 120, 8000 Aarhus C, Denmark}
\affiliation{Aarhus Institute of Advanced Studies, Aarhus University, H{\o}egh-Guldbergs Gade 6B, 8000 Aarhus C, Denmark}
\date{\today}

\begin{abstract}
	Quantum algorithms are promising candidates for the enhancement of computational efficiency for a variety of computational tasks, allowing for the numerical study of physical systems intractable to classical computers. In the Noisy Intermediate Scale Quantum (NISQ) era of quantum computing, however, quantum resources are limited and thus quantum algorithms utilizing such resources efficiently are highly coveted. We present a resource-efficient quantum algorithm for bosonic ground and excited state computations using the Variational Quantum Eigensolver algorithm with the Unitary Coupled Cluster ansatz. The algorithm is based on two quantum resource reduction strategies, consisting of a selective Hamming truncation of the encoded qubit Hilbert space along with a qubit ground state encoding protocol. Our algorithm proves to significantly increase accuracy with a simultaneous reduction of required quantum resources compared to current approaches. Furthermore, the selective Hamming truncation of our algorithm presents a versatile method to tailor the utilized quantum resources of a quantum computer depending on the hardware parameters. Finally, our work may contribute to shortening the route to achieve a practical quantum advantage in bosonic quantum simulations. The study of vibrational properties of molecular systems is crucial in a variety of contexts, such as spectroscopy, fluorescence, chemical reaction dynamics and transport properties. Thus, our algorithm provides a resource-efficient flexible approach to study such applications in the context of quantum computational chemistry on quantum computers.
\end{abstract}

\maketitle 

\section{Introduction}
The digital simulation of many-body physics is one of the most promising applications of quantum computers \cite{Lloyd1073}. Yet, current Noisy Intermediate Scale Quantum (NISQ) devices are faced with the challenges of limited quantum resources such as qubit memory and quantum gate fidelities \cite{Preskill_2018,2021arXiv210108448B}. For quantum algorithms to be feasible for such devices, it is crucial to consider algorithms which reduce the required resources for equivalent computations \cite{2018PhRvA..98b2322B,Peruzzo_2014,Kandala_2017,2020arXiv200805012D,2020Sci...369.1084.,PhysRevX.8.011021,https://doi.org/10.1002/qua.25176,2003SPIE.5105...96S,Sawaya2019}. A highly pursued application of quantum computing is quantum computational chemistry \cite{Aspuru-Guzik1704}, for which significant developments have been made in the numerical study of molecular systems using the Variational Quantum Eigensolver (VQE) algorithm \cite{Romero_2018,RevModPhys.92.015003,C9SC01313J,D0SC01908A,2020arXiv200806562F}. Specifically, the study of the vibrational properties of molecular systems is crucial in a variety of contexts, such as spectroscopy, fluorescence, chemical reaction dynamics and transport properties \cite{C9SC01313J,Crim12654,doi:10.1021/jp0370324,ZHU2009137,HUANG2012110}. With bosonic quantum simulations possibly providing a platform to achieve an early quantum advantage for practical tasks compared to its electronic counterpart, the development of resource-efficient bosonic quantum algorithms is crucial \cite{sawaya2020near}.\\
In the development of such algorithms, the encoding of bosonic degrees of freedom has received extensive attention in the literature with the optimal encoding protocol being highly sensitive to the structure of the bosonic operators to be implemented. Specifically, the concept of Hamming distance between the encoded qubit states has proved to be an important measure in this context \cite{D0SC01908A,sawaya2020near,Sawaya2019}. In current research, benchmark encoding protocols include the direct mapping (DM), standard binary (SB) and Gray code (GC) \cite{Sawaya2019,2005PhDT.......361S,PhysRevLett.121.110504,PhysRevA.98.042312,2003SPIE.5105...96S}. The DM encodes bosonic occupations to the states of individual qubits, with the encoded Hilbert space scaling as $\textrm{dim}(\mathcal{H}_\textrm{DM}) = N_\textrm{qubits}$. The SB and GC encode bosonic occupations into collective qubit states, achieving an exponential dimensional scaling compared to the DM with $\textrm{dim}(\mathcal{H}_\textrm{SB/GC}) = 2^{N_\textrm{qubits}}$. The differing amounts of bits between encoded qubit states, i.e. the amount of bit flips required to perform a transition from one qubit state to another, defined as the Hamming distance, is an important measure when transforming bosonic operators into a qubit representation. Thus, bosonic operators for which its qubit representation exhibits a high Hamming distance requires more operations to implement. As an illustrative example, a nearest-neighbour bosonic operator, as is present in bosonic systems such as the Bose-Hubbard model \cite{PhysRev.129.959,1986PhRvB..34.3136M}, may be efficiently implemented using GC. Such an efficient implementation is achieved since neighbouring qubit states in the GC, by definition, exhibit Hamming distances of one. While there exists no general rules for optimal encodings for general $d$-level Hamiltonians, certain trends may be observed as was demonstrated in Ref. \cite{Sawaya_2020}. In some instances, it may be beneficial to perform intermediate conversions between encodings to achieve optimal resource reductions. For other cases, some encodings allow for simultaneous reductions in both qubit and quantum gate resources.\\
In VQE approaches to quantum computational chemistry, the Unitary Coupled Cluster (UCC) ansatz has been widely applied in the context of fermionic simulations \cite{Peruzzo_2014,McClean_2016,2018arXiv181002327L,2017arXiv170102691R,Kandala2017}. Using a bosonic formulation of the UCC ansatz, bosonic ground and excited state computations using the VQE have been implemented using the DM in Ref. \cite{D0SC01908A}. To study impacts of different encodings, we investigate different encodings of a bosonic UCC ansatz and benchmark the DM, SB and GC encodings. We verify numerically that compact encodings for bosonic UCC exhibit infeasible quantum circuit depth scalings for implementation, as was suggested, although with no direct evidence, in Ref. \cite{D0SC01908A}. Scalings for direct encodings relative to compact encodings may be understood in terms of Hamming distances. In bosonic coupled cluster methods, the reference state amounts to a single-occupied Hartree product from which excitations are generated into unoccupied modal states using the cluster operator. Thus, the relevant bosonic excitation operators exhibit identical structures for each mode, namely $a^{\dagger}_{\text{unocc.}}a_{\textrm{occ.}}$, for which the same occupied modal state is annihilated for every coupled cluster excitation. In a qubit representation, such excitation operators exhibit identical Hamming distances of two, leading to a smooth scaling of circuit depths. In contrast, compact mappings yield differing Hamming distances depending on the encoded qubit states. Specifically, the circuit depths scale critically depending on the qubit register sizes due to increasing Hamming distances for the cluster excitation operators. As a result, due to the structure of the bosonic excitation operators, compact mappings generally imply larger circuit depths using coupled cluster methods. Thus, the typical advantages obtained using compact encodings such as the GC seem be to absent in bosonic UCC. Scaling with the mode truncations of the bosonic system, the cluster operator thus accumulates larger quantum circuit depths due to increasing Hamming distances with no feasible cancellations of quantum gates due to, for example, exploitation of commutation relations. Such a scaling is contrasted to direct encodings for which the excitation operators exhibit a constant Hamming distance.\\
To mitigate the large quantum circuit depths of compact encodings and the large qubit registers for the DM, allowing for compact bosonic VQE computations, we present a resource-efficient encoding algorithm, the Compact Encoding Algorithm (CEA), for the efficient computation of bosonic ground and excited state energies. Specifically, we propose a selective Hamming truncation of the encoded qubit Hilbert space along with a ground state encoding protocol, utilizing the natural structure of bosonic energy spectra. The selective Hamming truncation allows for systematic reductions of high Hamming distance excitations while the ground state encoding protocol eliminates redundant annihilations performed in direct mappings. Using these strategies, the CEA allows for a simultaneous reduction in both qubit and quantum gate resources, contributing to a shorter route to achieve a quantum advantage for bosonic quantum simulations. Since smaller qubit registers and shallower circuit depths give rise to much lower probabilities of errors, such reductions yield substantially increased accuracy while utilizing fewer quantum resources. To demonstrate the improved accuracy, we provide a numerical study of the common benchmark molecule, the CO$_{2}$ molecule, for which we perform ground and excited state computations using the VQE and the QEOM algorithm \cite{ollitrault2020quantum} both at current and future hardware parameters. Relative to the current approaches using the DM, the CEA provides a polynomial increase in qubit memory scaling with the Hamming truncation. With improving quantum gate fidelities, the Hamming truncation may be relaxed such that the CEA asymptotically approaches the exponential qubit memory of the SB. Thus, the CEA provides a flexible method to tailor the required quantum resources depending on the hardware parameters.\\
The manuscript is structured as follows. In Section \ref{sec:compact_encoding_algorithm}, we present the encoding strategies of the Compact Encoding Algorithm. In Section \ref{sec:comparison}, we present a comparative study of the CNOT gate counts required to implement the bosonic UCC operator using direct and compact encoding protocols to illustrate the different dimensional scalings. In Section \ref{sec:CO2}, we present a numerical study of ground and excited states of the CO$_{2}$ molecule using the CEA and the DM. Finally, in Section \ref{sec:conclusion}, we provide concluding remarks and outlooks.
\section{Compact encoding algorithm}
\label{sec:compact_encoding_algorithm}
In the following, we describe the two strategies of the CEA. In Section \ref{sec:SB_encoding}, the encoding of bosonic degrees of freedom using the SB protocol along with the associated qubit representation of the bosonic operators is presented. In Section \ref{sec:Hamming_truncation}, the selective Hamming truncation is introduced. Finally, in Section \ref{sec:GSEP}, the ground state encoding protocol (GSEP) is presented.
\subsection{Standard binary encoding of bosonic degrees of freedom}
\label{sec:SB_encoding}
Compared to the two-body Coulomb interactions of electronic systems, an interacting system of bosons exhibits many-mode couplings which further complicates the structure of the Hamiltonian. One way to parametrize a many-body system of interacting bosons is to expand the interaction term in a so-called $n$-body expansion \cite{doi:10.1063/1.1637578,10.1063/1.1637579}. In this expansion, each bosonic mode is expanded into a spectrum of modals for which the total state of the system is described as a product state of occupation number vectors designating the modal occupations for each mode, called a Hartree product. This formalism and the UCC ansatz are used in the following sections for which additional details may be found in Appendix \ref{appendix:bosonic_many_body_Hamiltonian} and \ref{appendix:UCC}, respectively.\\
Consider a bosonic system with $L$ modes with modal dimension $N_{l}$ for mode $l$. A configuration for such a system may be parametrized by a configuration vector, $\textbf{r} = [r^{0},...,r^{L-1}]^{T}$, with $\{r^{l}\}$ designating the indices of the occupied modals. The corresponding Hartree product reads
\begin{align}
\ket{\Phi_{\textbf{r}}} &= \ket{r^{0}}\otimes...\otimes\ket{r^{L-1}}\\
&=\underbrace{\Big(\bigotimes_{i=0}^{\log_{2}(N_{0})-1}\ket{r^{0}_{i}}\Big)}_{\text{Mode 0 register}}\otimes...\otimes\underbrace{\Big(\bigotimes_{i=0}^{\log_{2}(N_{L-1})-1}\ket{r^{L-1}_{i}}\Big)}_{\text{Mode L-1 register}}
\end{align}
where $r^{l}_{i}\in\{0,1\}$ is the $i$'th coefficient in the binary decomposition of $r^{l}$.\\
The mapping of the annihilation operator is defined as
\begin{equation}
a_{r^{l}}^{l}\rightarrow\bigotimes_{i=0}^{\log_{2}(N_{l})-1}\Big(\sigma_{i}^{l -}\Big)^{r^{l}_{i}}
\label{eq:annihilation_definition}
\end{equation}
where the Pauli operator acts on the $i$'th qubit in the $l$'th register. The analogous definition holds for the Hermitian conjugate of Eq. \ref{eq:annihilation_definition}.
\\\\
\fbox{\begin{minipage}{\columnwidth}
		\begin{center}
			\subsubsection*{Hamming distance}
		\end{center}
		For two binary strings, the Hamming distance is defined as the amount of bits differing between the two strings. As an example, consider the strings 001 and 101. For these two strings, $\textrm{dh}(\textrm{001}, \textrm{101}) = 1$ since the first bit entry differs.
\end{minipage}}\\
\subsection{Strategy 1: Hamming truncation}
\label{sec:Hamming_truncation}
When implemented on a quantum computer, the UCC ansatz is decomposed in a Trotter expansion which allows for the sequential operation of quantum logic gates on the reference Hartree product. Using Eq. \ref{eq:annihilation_definition}, excitation operators for the UCC transform as
\begin{equation}
a_{r^{l}}^{l \dagger}a_{s^{l}}^{l}\rightarrow\bigotimes_{i=0}^{\log_{2}(N_{l})-1}\Big(\sigma_{i}^{l +}\Big)^{r^{l}_{i}}\Big(\sigma_{i}^{l -}\Big)^{s^{l}_{i}}.
\label{eq:excitation_operator}
\end{equation}
Using the DM, all encoded qubit states differ by two qubits, yielding a constant Hamming distance of $\textrm{dh} = 2$ for all states. With reference to Eq. \ref{eq:excitation_operator}, a cluster excitation operator in the DM exhibits the same Pauli structure for all excitations. This is generally not true for compact encodings, since compact encodings contain states with $\textrm{dh} > 2$ scaling with the qubit register sizes for each mode. For $\textrm{dh} = 4$, according to Eq. \ref{eq:excitation_operator}, this yields a product of four Pauli raising/lowering operators. Subtracting the Hermitian conjugate yields a total of 8 terms of Pauli operators with 4 products in each term. To avoid lengthy Pauli terms, one may selectively truncate the Hilbert space to contain only qubit states with a Hamming distance below a given threshold. Thus, one may decompose the qubit Hilbert space into two subspaces,
\begin{equation}
\mathcal{H} = \mathcal{H}_{\textrm{enc}}\cap\mathcal{H}_{\textrm{trun}}
\end{equation}
where $\mathcal{H}_{\textrm{enc}}$ is the subspace of qubit states available for encoding and $\mathcal{H}_{\textrm{trun}}$ is the truncated subspace. The result of the Hamming truncation produces $\mathcal{H}_{\textrm{enc}}$ which requires a shallower circuit depth for implementation.\\
By performing a Hamming truncation of compactly encoded qubit states, one does not achieve an exponential qubit memory advantage as would be obtained using the SB/GC. However, scaling with the Hamming truncation, the CEA provides a polynomial qubit memory increase compared to the DM. With progressive improvements of quantum hardware, the Hamming truncation may be relaxed to include higher Hamming distance states, asymptotically approaching the exponential qubit memory scaling of the SB/GC. Thus, for a given set of available qubit resources, the CEA exhibits a polynomial qubit memory advantage as compared to the DM. The binary decomposition of the modal indices, however, no longer necessarily correspond to the qubit states. One may therefore transform the modal states into, in principle, arbitrary qubit states.
\subsection{Strategy 2: Ground state encoding protocol}
\label{sec:GSEP}
Using the DM, one must systematically annihilate the reference occupations of the reference Hartree product. Such redundant operations may be eliminated using compact encodings by encoding the reference modal states into the qubit ground states. This is achieved by re-organizing the modal basis elements for each mode such that the modals contained in the reference configuration obtain the label $s^{l} = 0$. With this re-ordering, all reference modals are transformed into their respective qubit register ground states, $\ket{0...0}$. This has no physical implications since the ordering of basis elements is arbitrary.\\
The excitation operators then reduce significantly with
\begin{equation}
\bigotimes_{i=0}^{\log_{2}(N_{l})-1}\Big(\sigma_{i}^{l +}\Big)^{r^{l}_{i}}\Big(\sigma_{i}^{l -}\Big)^{s^{l}_{i}} = 	\bigotimes_{i=0}^{\log_{2}(N_{l})-1}\Big(\sigma_{i}^{l +}\Big)^{r^{l}_{i}}.
\end{equation}
This is the case since $s^{l}_{i}=0$ for all $i$ and all $l$ in the reference Hartree product. Thus, all annihilation operators equal the identity operator, yielding shallower quantum circuit depths.
\section{Comparison of bosonic encoding protocols}
\label{sec:comparison}
\begin{figure*}
	\centering
	\includegraphics[width=1.5\columnwidth]{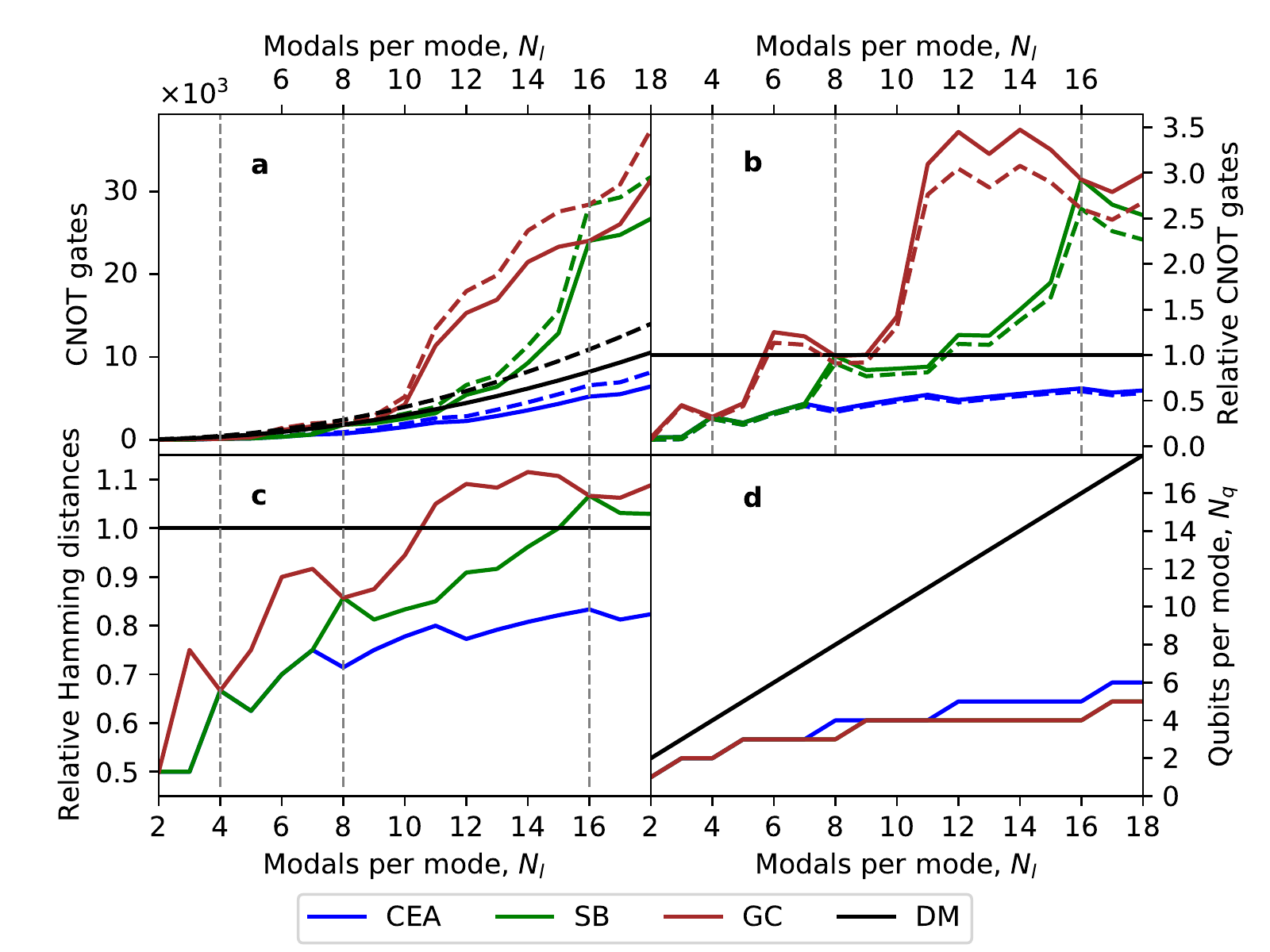}
	\caption{Illustration of quantum resource scalings for the different encoding protocols, the CEA, SB, GC and DM. The bosonic UCC quantum circuits were generated for two bosonic modes for different modal space dimensions. Subsequently, the quantum circuits were optimized, taking into account the cancellation of quantum gates based on commutation relations. The bosonic UCC operator was constructed using single and double excitation operators (UCCSD). \textbf{a)} The required number of CNOT gates for the implementation of the bosonic UCC operator as a function of the modal dimension per mode, $N_{l}$. The dashed lines correspond to non-optimized quantum circuits while the solid lines correspond to optimized quantum circuits. The vertical dotted lines highlight the modal dimensions for which $\mathcal{H}_{\textrm{enc}}^{\textrm{SB}}=\mathcal{H}_{\textrm{enc}}^{\textrm{GC}}$. Note the systematic reduction in CNOT gates for the CEA as a result of the ground state encoding protocol and selective Hamming truncation. \textbf{b)} Relative number of required CNOT gates required for implementation relative to the DM as a function of the modal dimension per mode, $N_{l}$. \textbf{c)} The amount of total Hamming distances of the qubit cluster excitation operators for the bosonic UCC operators relative to the DM as a function of the modal dimension per mode, $N_{l}$. \textbf{d)} Amount of qubits per mode as a function of the modal dimension per mode, $N_{l}$. The DM exhibits a linear dependence on qubit register sizes as a function of $N_{l}$. The SB and GC coincide for all modal dimensions since the two encodings differ only in the ordering of basis elements. The CEA coincides with the SB/GC for small dimensions and mitigates high Hamming distance states by the encoding of additional qubits (see text for details).}
	\label{fig:comparison}
\end{figure*}
In the following, we present a comparison of the quantum resource scalings for the four different mappings, namely the CEA, SB, and GC based on different compact encoding strategies and the direct encoding, the DM. Since CNOT gates provide a significant overhead in quantum gate resources compared to one-qubit gates, we benchmark the encoding protocols based on their required CNOT gates for implementation. Specifically, the bosonic UCC quantum circuits were generated for two bosonic modes for different modal space dimensions. The bosonic UCC operator was constructed using single and double excitation operators (UCCSD). Subsequently, the quantum circuits were optimized using the quantum circuit transpiler method in Qiskit with maximum optimization, taking into account the cancellation of quantum gates based on commutation relations. The number of CNOT gates required for implementation as a function of the modal space dimensions per mode, $N_{l}$, are presented in Fig. \ref{fig:comparison}a. Relative CNOT gate counts are presented relative to the DM in Fig. \ref{fig:comparison}b. The total amount of relative Hamming distances (RHD) for all single and double cluster excitation operators in the UCC operator, relative to the DM, are presented in Fig. \ref{fig:comparison}c. Finally, the required qubit register dimensions, $N_{q}$, are presented in Fig. \ref{fig:comparison}d. Using these figures, three conclusions may be drawn.\\
First, as would be expected, the RHD appears to effect the amount of CNOT gates for implementation, as is evident from Figs. \ref{fig:comparison}a, \ref{fig:comparison}b and \ref{fig:comparison}c. Increases in the RHD appear to be correlated with increases in the CNOT gates. The GC encodes high Hamming distance states at smaller dimensions compared to the SB to maintain nearest-neighbour Hamming distances of one, reflected in the rapid increase of the RHD in Fig. \ref{fig:comparison}c. Due to the structure of the bosonic cluster excitation operators, however, it appears that the GC does not provide any advantage compared to the SB since no feasible cancellations and nearest-neighbour interactions may be exploited. In fact, the GC appears to accumulate RHD and CNOT gates more rapidly as compared to the SB. As would be expected, the RHD and required CNOT gates for the SB and GC intersect at dimensions $\textrm{dim}(\mathcal{H}^{\textrm{SB/GC}}_\textrm{enc}) = 2^{N_\textrm{q}}$ since $\mathcal{H}_{\textrm{enc}}^{\textrm{SB}}=\mathcal{H}_{\textrm{enc}}^{\textrm{GC}}$, as can be seen at $N_{l}=4,8,16$. This is the case since the encoded qubit Hilbert spaces are equal at these dimensions, differing only in the ordering of basis elements. As a result, for these dimensions, the bosonic UCC operators for the SB and the GC contain identical qubit cluster excitation operators, yielding equal RHD and CNOT gates.\\
Second, for $N_{l}<6$, corresponding to $N_{q}\leq3$, the GSEP provides advantages to all compact encodings since no ground state annihilations are performed as compared to the DM, systematically reducing the required CNOT gates. At $N_{l}=6$, the GC encodes $\ket{111}$ with $\textrm{dh}=3$ in contrast to the constant $\textrm{dh}_{\textrm{DM}}=2$, increasing the relative amount of CNOT gates. Without the GSEP, all excitations in the SB and the GC would yield qubit excitation operators for which $\textrm{dh}\geq2$, systematically requiring more CNOT gates for implementation relative to the DM.\\
Third, at $N_{l} = 11$ ($N_{l} = 16$), the GC (SB) no longer benefits from the GSEP since the CNOT gates surpass the DM due to rapidly increasing RHD. Such rapid increases in RHD are results of the encoding of high Hamming distance states due to increasing qubit register sizes. The SB and CEA exhibit equal RHD and CNOT gates until $N_{l}=8$ at which the SB encodes $\ket{111}$. The encoding of the high Hamming distance states, and the associated increase in CNOT gates, may be mitigated by the addition of a qubit to the register. This expands the selective truncation and thus the encoded subspace, allowing for the encoding of inexpensive qubit states. This procedure amounts to the most important concept of the CEA. As can be seen in Fig. \ref{fig:comparison}, for $N_{l}\geq8$, by selectively excluding high Hamming distance qubit states with $\textrm{dh}\geq3$, the CEA provides a systematic reduction of RHD and CNOT gates. Such reductions are compensated by increasing $N_{q}$, as is illustrated in Fig. \ref{fig:comparison}c. For NISQ devices, however, the trade-off is feasible since the CEA maintains smaller qubit registers compared to the DM with a simultaneous reduction of quantum gate resources. With improving CNOT gate fidelities, however, one may relax the Hamming truncation to include states with higher Hamming distances, asymptotically approaching the exponential qubit memory advantage of the SB/GC.\\
The results discussed in this section are generic and show clearly the advantages and disadvantages of each of the encoding protocols. In particular, we conclude that using the two strategies of the CEA, i.e. the ground state encoding protocol and selective Hamming truncation, compact encodings suitable for NISQ-era devices may be achieved.
\section{Numerical study of $\textrm{CO}_{2}$}
\label{sec:CO2}
\begin{figure*}
	\centering
	\includegraphics[width=1.5\columnwidth]{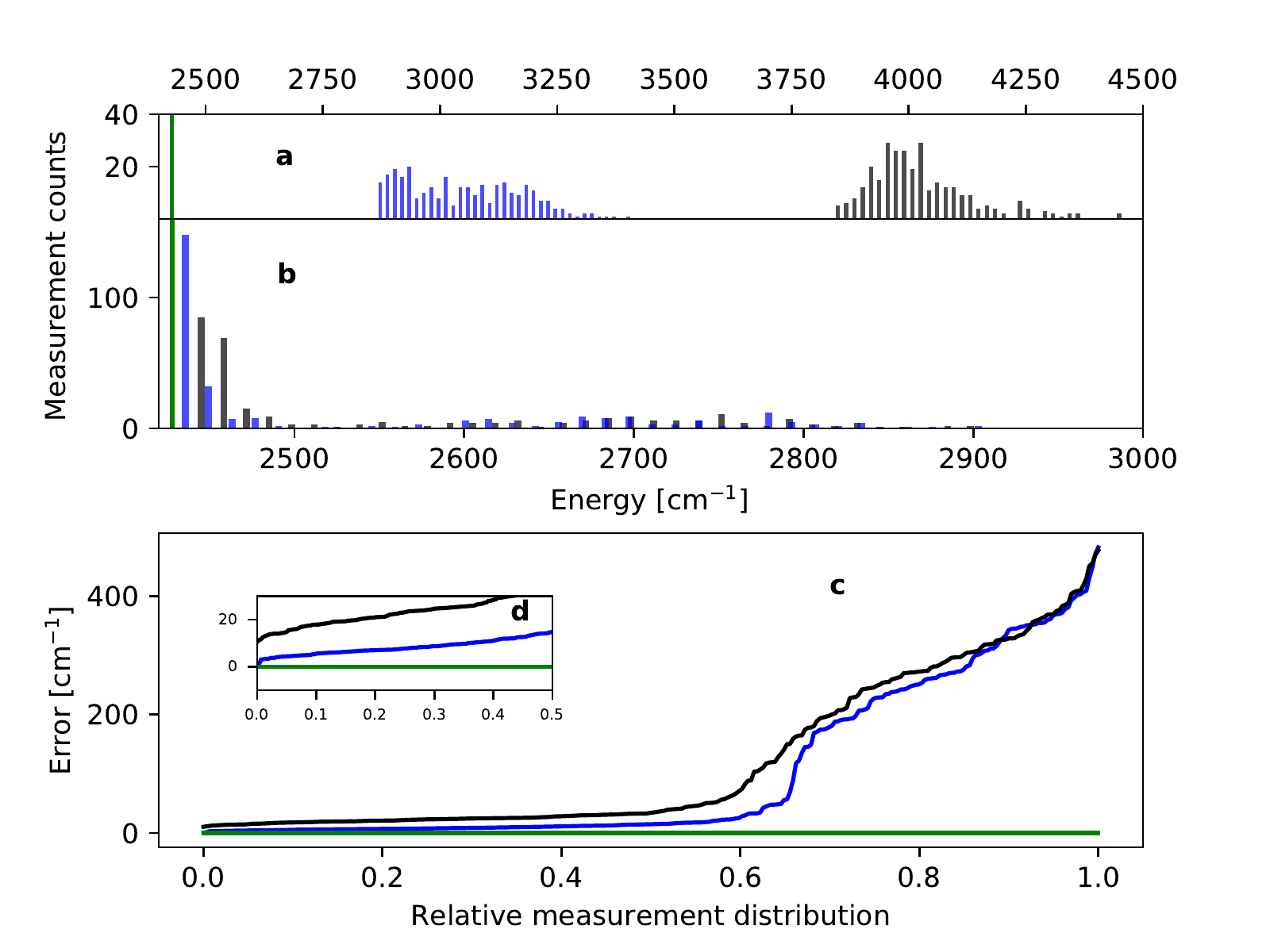}
	\caption{Ground state energy calculations for the CEA and the DM with 300 VQE samples for each algorithm. Blue lines represent CEA measurements, black lines represent DM measurements and the green line represents the reference energy. Bin sizes of measurement samples are 35. Note the different axis scalings. \textbf{a)} Histogram of sampled VQE measurements with hardware parameters $(1.0\times10^{-3}, 1.0\times10^{-2})$. With current hardware parameters, the CEA provides a large increase of accuracy with an improvement of nearly 1000 cm$^{-1}$. \textbf{b)} Histogram of sampled VQE measurements with hardware parameters $(0.5\times10^{-5}, 0.5\times10^{-4})$ at convergence. The subset of minimal energy measurements increases significantly for both algorithms with improving hardware parameters, which is advantageous in terms of the variational principle. \textbf{c)} Cumulative distribution of VQE measurement errors with hardware parameters $(0.5\times10^{-5}, 0.5\times10^{-4})$. The error is calculated relative to the reference energy. \textbf{d)} Zoom of \textbf{c} with the convergence of CEA highlighted. Note the large subset of measurements which exhibit minimal error compared to the reference energy.}
	\label{fig:ground_state}
\end{figure*}
\begin{figure}
	\centering
	\includegraphics[width=1.0\columnwidth]{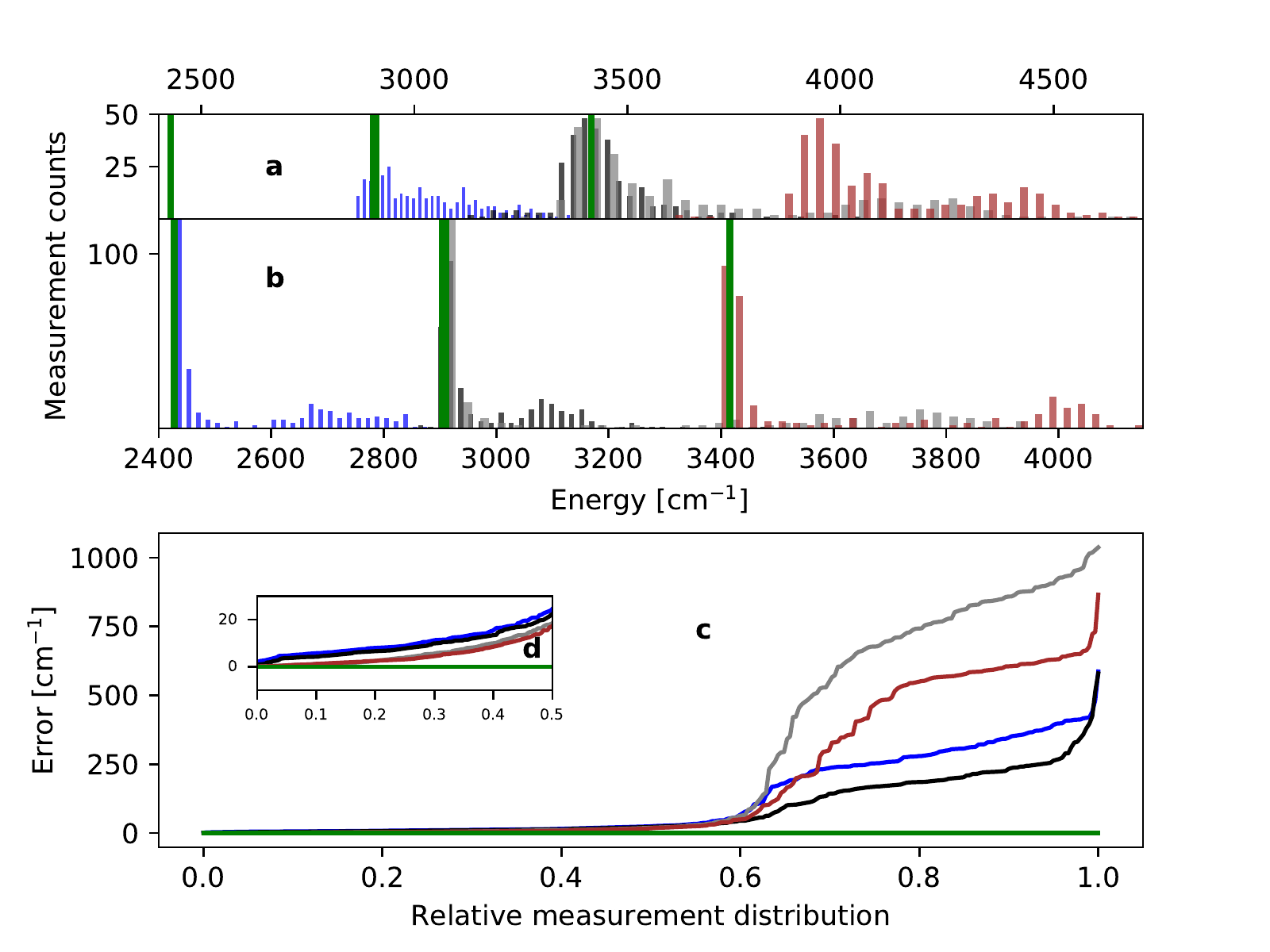}
	\caption{Excited state energies for the CEA with 300 VQE samples. The lines represent the excited states as (blue, ground state), (black, first excited state), (gray, second excited state), (brown, third excited state) and the green lines represent the reference energies. The first and second excited states are degenerate.  Bin sizes of measurement samples are 35. Note the different axis scalings. \textbf{a)} Histogram of sampled VQE measurements with hardware parameters $(1.0\times10^{-3}, 1.0\times10^{-2})$. \textbf{b)} Histogram of sampled VQE measurements with hardware parameters $(0.5\times10^{-5}, 0.5\times10^{-4})$. With improving hardware parameters, the excited state energies calculations converge to the reference energies using the QEOM algorithm analogously to the ground state calculations. \textbf{c)} Cumulative distribution of VQE measurement errors with hardware parameters $(0.5\times10^{-5}, 0.5\times10^{-4})$. The errors are calculated relative to the reference energies. \textbf{d)} Zoom of \textbf{c} with the convergence of CEA highlighted. Note the large subset of measurements which exhibit minimal error compared to the reference energies.}
	\label{fig:CEA_excited_state}
\end{figure}
\begin{figure}
	\centering
	\includegraphics[width=1.0\columnwidth]{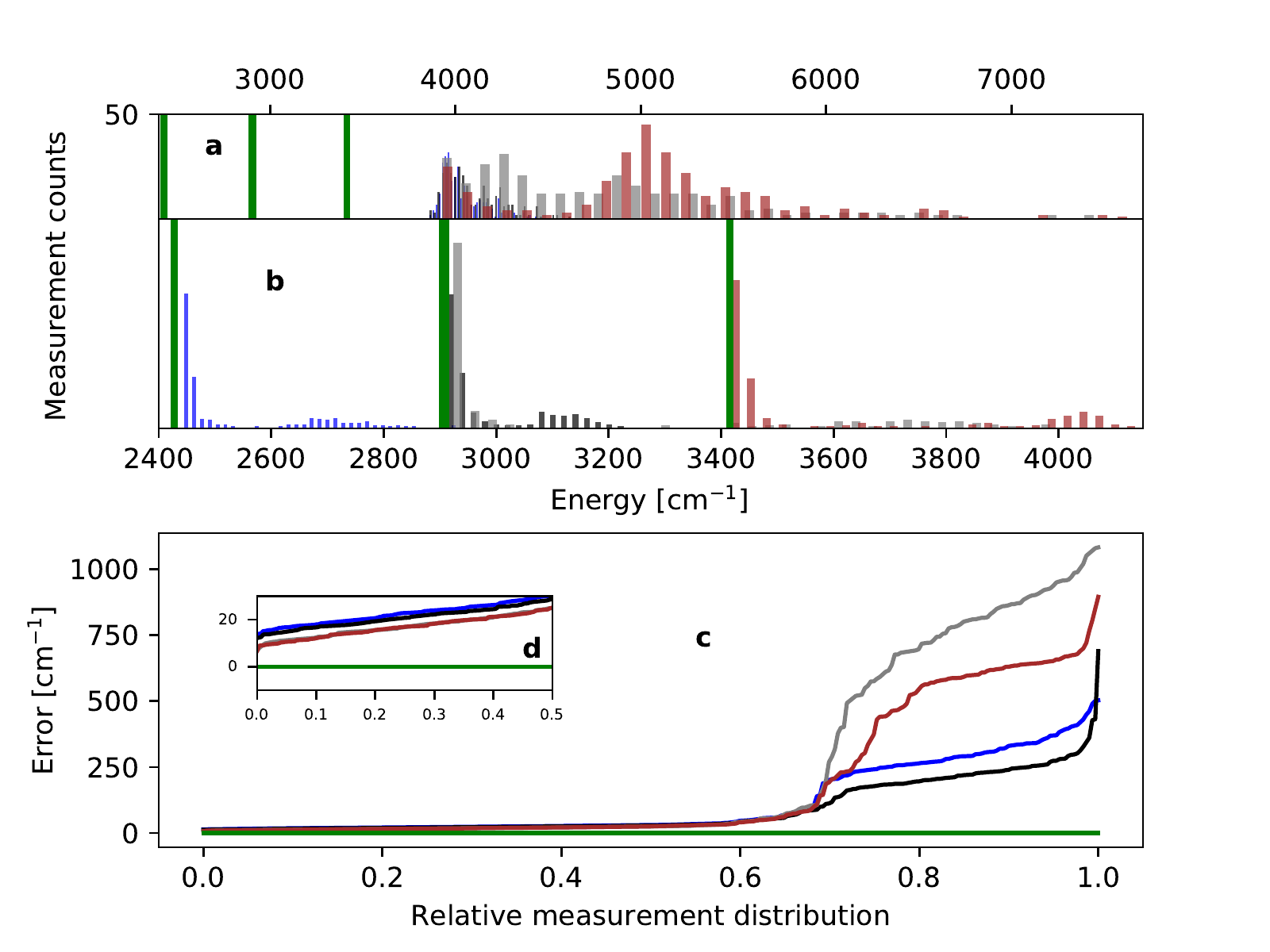}
	\caption{Excited state energies for the DM with 300 VQE samples. The lines represent the excited states as (blue, ground state), (black, first excited state), (gray, second excited state), (brown, third excited state) and the green lines represent the reference energies. The first and second excited states are degenerate.  Bin sizes of measurement samples are 35. Note the different axis scalings. \textbf{a)} Histogram of sampled VQE measurements with hardware parameters $(1.0\times10^{-3}, 1.0\times10^{-2})$. \textbf{b)} Histogram of sampled VQE measurements with hardware parameters $(0.5\times10^{-5}, 0.5\times10^{-4})$. With improving hardware parameters, the excited state energies calculations approach the reference energies, although without convergence as was observed using the CEA. \textbf{c)} Cumulative distribution of VQE measurement errors with hardware parameters $(0.5\times10^{-5}, 0.5\times10^{-4})$. The errors are calculated relative to the reference energies. \textbf{d)} Zoom of \textbf{c}.}
	\label{fig:DM_excited_state}
\end{figure}
In the following, we present illustrative computational studies of the ground and excited state energies for the common benchmark molecule, $\textrm{CO}_{2}$. In order to demonstrate the significant increase of precision using the CEA, the computational studies are benchmarked relative to current approaches using the DM \cite{D0SC01908A}. For $\textrm{CO}_{2}$, we calculate the ground state energy along with three excited state energies using the Quantum Equation of Motion (QEOM) algorithm \cite{PhysRevResearch.2.043140} and compare the results to reference energies (exact diagonalization of the Hamiltonian). All VQE computations are simulated with the noisy device QASM simulator using an extended version of the IBM Qiskit software package \cite{Qiskit} implementing the CEA. All four vibrational modes of $\textrm{CO}_{2}$ are included with two modal wavefunctions per mode, requiring four qubits for the CEA and eight qubits for the DM. We sample ground and excited state energies with 300 samples for both the CEA and the DM. Further computational details are presented in Appendix \ref{appendix:numerical_simulation_details}. To investigate the impact of hardware improvements and to study the comparative performance at convergence, we vary the hardware parameters until convergence is achieved for the $\textrm{CO}_{2}$ computations. With the 2-qubit gate fidelity improving one order of magnitude, we assume the one-qubit gate fidelities to also improve one order of magnitude. In the following, we represent hardware parameters in the format (prob. for 1-qubit error, prob. for 2-qubit error).
\subsection{Computation of ground state energies}
In Fig. \ref{fig:ground_state}a, we present a histogram of the sampled VQE computations with current hardware parameters. While both algorithms exhibit large errors compared to the reference energy, the CEA demonstrates a significant improvement of accuracy. Using current hardware parameters estimated to be $(1.0\times10^{-3}, 1.0\times10^{-2})$ as reference points, we varied the hardware parameters until convergence was achieved, as depicted in Fig. \ref{fig:ground_state}d. Specifically, we achieve convergence at hardware parameters $(0.5\times10^{-5}, 0.5\times10^{-4})$. Note that the measured energy functional in the VQE algorithm, $E[\mathbf{\theta}]$, is optimized according to the variational principle, yielding a guaranteed upper bound to the exact ground state energy. Therefore, the relevant subset of measurements are those which are minimal. For hardware parameters at convergence, a distribution of measurement counts and a corresponding cumulative error distribution are presented in Figs. \ref{fig:ground_state}b and \ref{fig:ground_state}c. Results for both algorithms appear to accumulate at minimal energies. For both the CEA and the DM, there is a clear tendency to produce large subsets of minimal energy measurements which is both evident in the histogram and cumulative distribution of Figs. \ref{fig:ground_state}b and \ref{fig:ground_state}c. While both algorithms seem to produce relatively accurate results in Fig. \ref{fig:ground_state}d, with almost half of all measurements within an accuracy of $<20$ cm$^{-1}$, the CEA consistently achieves a higher accuracy and, at the same time, requires fewer quantum resources.
\subsection{Computation of excited state energies}
The results of the QEOM computations for the excited state energies are presented for the CEA and the DM in Figs. \ref{fig:CEA_excited_state}  and \ref{fig:DM_excited_state} for which identical comparative conclusions as above may be made.
\section{Conclusion}
\label{sec:conclusion}
Compact encoding strategies for bosonic UCC methods are believed to be inferior to direct mappings due to quantum circuits depths that scale unfavorably for near-term noisy quantum hardware. As we have demonstrated in our numerical studies, the standard binary and the Gray code do indeed require large amounts of quantum gate resources compared to direct encodings. We remedy this unfortunate situation by proposing a new encoding algorithm, the Compact Encoding Algorithm (CEA), for bosonic quantum simulations using variational approaches. The CEA overcomes the drawbacks of current approaches to compact encodings by using a ground state encoding protocol and a selective Hamming truncation. This allows for a simultaneous reduction in both qubit and quantum gate resources compared to all of the previous approaches. Therefore, in the short-term perspective, with a major increase of precision while utilizing fewer quantum resources, we expect the algorithm to contribute to shortening the route to a practical quantum advantage for bosonic quantum simulations. Furthermore, the reduction of quantum resources automatically allows for the study of larger molecular systems with current hardware. Finally, the strategies of our work allow for a flexible method to tailor the required quantum resources depending on the hardware parameters of the quantum computer.\\
While the Hamming truncation specifically targets NISQ era quantum devices, we expect the Hamming truncation to provide encoding flexibility for future fault-tolerant quantum devices by appropriately choosing the Hamming threshold based on the preferences of the hardware parameters. Specifically, for quantum devices favoring larger (smaller) qubit registers with shallower (deeper) quantum circuit depths, one may choose Hamming distances which feasibly allocate the required quantum resources depending on the hardware parameters.
\appendix
\section{Bosonic many-body Hamiltonian in the $n$-body expansion formalism}
\label{appendix:bosonic_many_body_Hamiltonian}
A many-body system of interacting bosons, in general, exhibit rather complicated interactions compared to, for example, electronic interactions in molecules. With electrons interacting through the two-body Coulomb interaction, bosons may interact through many-body coupling terms which further complicates the structure of the Hamiltonian. One way to parametrize a many-body system of interacting bosons is to expand the interaction term in a so-called $n$-body expansion. With the $n$-body expansion, one may obtain an accurate description of the anharmonicity of the potential, thus avoiding the intricate structure of Taylor expansions. The degrees of freedom of the bosonic system may be parametrized by a set of normal modes of vibrations, $L$. Each mode may be parametrized by a normal coordinate, $q_{l}$. The Hamiltonian reads
\begin{equation}
H = \sum_{l=0}^{L-1}\Big(-\frac{1}{2}\frac{\partial^2}{\partial q_{l}^{2}}\Big) + V(\{q_{l}\}).
\label{eq:Hamiltonian}
\end{equation}
Following Christiansen \cite{10.1063/1.1637579} and Ollitrault \cite{D0SC01908A}, the potential in the $n$-body expansion may be written as
\begin{equation}
V(\{q_{l}\}) = V_{0} + \sum_{l=0}^{L-1}V^{[l]}(q_{l}) + 
\sum_{l<m}^{L-1}V^{[lm]}(q_{l},q_{m}) + ...
\label{eq:potential}
\end{equation}
In this expansion, the first term contains the equilibrium configuration energy and the latter sums represent higher-order mode couplings. For the first sum, $q_{l}$ is varied and all other coordinates remain in their equilibrium configuration. For the second sum, $q_{l}$ and $q_{m}$ are varied with all other coordinates in equilibrium, i.e. introducing two-body couplings. The expansion of Eq. \ref{eq:potential} may be truncated to a given order $n$. One may note that the Hamiltonian exhibits a variety of symmetries. These symmetries include conservation of modal excitations in the sense that all vibrational modes contain only one excitation of a modal at any time. Furthermore, such a modal is confined to its mode. This gives rise to a mode-conserving excitation manifold.\\
The spectrum of a given mode $l$ may be divided into a subspace of $N_{l}$ modals. The modal basis set for mode $l$ is given by a set of functions \cite{doi:10.1063/1.1637578}
\begin{equation}
S_{l} = \{\phi_{0}^{(l)}(q_{l}),...,\phi_{N_{l}-1}^{(l)}(q_{l})\}.
\label{eq:modal_basis_set}
\end{equation}
The total wavefunction of the system will be a linear combination of all possible product wavefunctions of modals from each mode, called Hartree products,
\begin{equation}
\ket{\psi} = \sum_{k_{0}}^{N_0}...\sum_{k_{L-1}}^{N_{L-1}}C_{k_{0}...k_{L-1}}\phi_{k_{0}}^{(0)}(q_{1})...\phi_{k_{L-1}}^{(L)}(q_{L-1}).
\end{equation}
In the second quantization formalism, a Hartree product is given by
\begin{equation}
\ket{\Phi_{\textbf{r}}} = \prod_{l=0} ^{L-1}a_{r^{l}}^{l \dagger}\ket{0}
\label{eq:hartree_product}
\end{equation}
where $\textbf{r} = [r^{0},...,r^{L-1}]^{T}$ is a vector designating the indices of the occupied modals in the product. Thus, a given modal occupation may be represented by an integer in the interval $r^{l}\in\{0,...,N_{l}-1\}$. Conventionally, the modal functions are mean field solutions to the Hamiltonian and may be solved using, for example, vibrational self consistent field (VSCF) methods \cite{10.1063/1.1637579}.	
\section{Unitary coupled cluster theory}
\label{appendix:UCC}
The UCC ansatz reads
\begin{equation}
\ket{\textrm{UCC}} = e^{T - T^{\dagger}}\ket{\Phi_{\textbf{s}}}
\label{eq:ucc}
\end{equation}
where $T$ is the cluster operator and $\ket{\Phi_{\textbf{s}}}$ is a reference Hartree product with a reference modal configuration given by $\textbf{s}$. Typically, only single and double excitations of the above ansatz are sufficient for accurate calculations. The single and double cluster operators read
\begin{equation}
T_{1} = \sum_{l=0}^{L-1}\sum_{r^{l}}t_{r^{l}}^{l}a_{r^{l}}^{l \dagger}a_{s^{l}}^{l}
\end{equation}
and
\begin{equation}
T_{2} = \sum_{l<m}\sum_{r^{l}}\sum_{p^{m}}t_{r^{l}p^{m}}^{lm}a_{r^{l}}^{l \dagger}a_{p^{m}}^{m \dagger}a_{s^{l}}^{l}a_{q^{m}}^{m}
\end{equation}
with $r^{l}, p^{m}$ and $s^{l}, q^{m}$ being unoccupied and occupied modals, respectively, and $t^{l}_{r^{l}},t_{r^{l}p^{m}}^{lm}\in\mathds{R}$. Thus, the UCC ansatz generates a linear combination of single (double) excitations out of the reference Hartree product while confining each modal excitation within the same mode.
\section{Numerical simulation details}
\label{appendix:numerical_simulation_details}
All VQE simulations were performed using the IBM Qiskit software package \cite{Qiskit}. Since the CEA is not implemented in Qiskit, an extended version of Qiskit was developed and may be obtained on request to the authors. To simulate a noisy device, we used the Qiskit QASM simulator. With four vibrational modes and two modal wavefunctions per mode, we allocated four qubit registers of size one (4 qubits) for the CEA and four qubit registers of size two (8 qubits) for the DM. For the optimization in the VQE, we used the COBYLA optimization routine. The ground state optimization of $\textrm{CO}_{2}$ was performed using density functional theory (B3LYP) with the 6-31g basis set in Gaussian16 \cite{g16} using the Qiskit interface. The electronic potential energy surface (PES) was constructed using the $\verb|GaussianForcesDriver|$ in Qiskit \cite{D0SC01908A}. This involves the approximation of the PES as a quartic force field with semi-numerical differentiation of the analytical Hessian. In the computation of the matrix elements of the $n$-body expansion Hamiltonian of Eq. \ref{eq:Hamiltonian}, a harmonic oscillator basis set was used. Since the subject of this work is to study the relative performance of the CEA and the DM, the particular choice of basis set is not important and the results found here apply for other basis sets as well.
\section*{Acknowledgments}
We are grateful to Ove Christiansen, Pauline Ollitrault and Ivano Tavernelli for helpful and insightful discussions in relation to this work. The numerical results presented in this work were obtained at the Centre for Scientific Computing, Aarhus, http://phys.au.dk/forskning/cscaa/. The authors acknowledge support from the Independent Research Fund Denmark, the Carlsberg Foundation, and the Aarhus University Research Foundation.

\bibliographystyle{ieeetr}
\bibliography{ManuscriptBibTex}

\end{document}